\documentclass [aps,prb,onecolumn,amssymb,amsmath,showpacs,floatfix,superscriptaddress]{revtex4-2}

\def\XXint#1#2#3{{\setbox0=\hbox{$#1{#2#3}{\int}$}
     \vcenter{\hbox{$#2#3$}}\kern-.5\wd0}}

\usepackage{amsmath}
\usepackage{amssymb}
\usepackage{amsfonts}
\usepackage{graphicx}
\usepackage{verbatim}
\usepackage{bm}
\usepackage{mathbbol}
\usepackage[dvipsnames]{xcolor}
\setcitestyle{compress}
\usepackage{mathtools}
\usepackage{esint}
\usepackage{tabularx}
\usepackage[normalem]{ulem}
\usepackage{graphicx,epsfig,amsfonts,amssymb}
\usepackage{bm}
\usepackage{times}
\usepackage{lipsum}
\usepackage{verbatim}
\usepackage{hyperref}
\hypersetup{
    colorlinks,
    citecolor=blue,
    filecolor=blue,
	    linkcolor=blue,
    urlcolor=blue
}
\usepackage{comment}

\newcommand{\beg}{\begin{equation}}
\newcommand{\en}{\end{equation}}

\newcommand{\overbar}[1]{\mkern 1.5mu\overline{\mkern-1.5mu#1\mkern-1.5mu}\mkern 1.5mu}
 
\begin{document}

\title{Thermalization of Weakly Nonintegrable FPUT and Toda Dynamics: A Lyapunov Spectrum Perspective}

\author{Aniket Patra}
\affiliation{Center for Theoretical Physics of Complex Systems, Institute for Basic Science, Daejeon 34126, Republic of Korea}
\affiliation{Department of Physics, Pukyong National University, Busan 48513, Republic of Korea}

\author{Sergej Flach}
\affiliation{Center for Theoretical Physics of Complex Systems, Institute for Basic Science, Daejeon 34126, Republic of Korea}
\affiliation{Basic Science Program, Korea University of Science and Technology (UST), Daejeon, 34113, Republic of Korea}
\affiliation{Centre for Theoretical Chemistry and Physics, The New Zealand Institute for Advanced
Study (NZIAS), Massey University Albany, Auckland 0745,
New Zealand}

\begin{abstract}
    We study the thermalization slowing down of Fermi-Past-Ulam-Tsingou (FPUT) chains and of Toda chains with nonintegrable boundaries. 
    We focus on the transition from FPUT to harmonic chains, from FPUT to Toda chains with fixed boundaries, and from nonintegrable open boundary Toda to integrable fixed boundary Toda.
    We compute the Lyapunov spectrum and analyze its scaling properties upon approaching integrable limits.
    We analyze the scaling of the largest Laypunov exponent, the rescaled Lyapunov spectrum, and the Kolmogorov-Sinai entropy.
    Using additional analytic arguments we demonstrate evidence that all three cases are operating in the regime of a Long Range Network of nonintegrable perturbations.
\end{abstract}

\maketitle

\section{Introduction}

Conducted more than seventy years ago, the seminal numerical experiment of Fermi, Pasta, Ulam, and Tsingou (FPUT) on thermalization in oscillator chains with a quadratic-plus-cubic interaction potential is widely regarded as the first computational investigation of nonequilibrium many-body dynamics \cite{FPUT}. Their simulations showed that when the system is initialized in a low-frequency normal mode of the corresponding linear chain, characterized by frequency $\omega_q$ and wavenumber $q$, the energy remains confined to a small set of neighboring modes for unexpectedly long times. Equipartition in mode space is eventually achieved, but only on exceptionally long timescales. Remarkably, despite the dynamics being governed by a nonlinear Hamiltonian that exhibits chaotic behavior, the system displays near-recurrences to the initially excited mode before thermalizing.

These observations stood in sharp contrast to the expectations of statistical mechanics and were initially perceived as paradoxical. The subsequent effort to understand this paradox has driven major developments in modern statistical and mathematical physics \cite{FPUT_Ford, FPUT_Weissert, FPUT_Gallavotti, FPUT_Chaos1, FPUT_Chaos2}. Indeed, a satisfactory explanation of the FPUT results requires the combined application of powerful theoretical frameworks -- including Kolmogorov-Arnol'd-Moser (KAM) theory \cite{KAM11, KAM12, KAM13, KAM3, KAM4, KAM_Perturb1, KAM_Perturb2, KAM_Perturb3, FPUT_KAM_1, FPUT_KAM_2, FPUT_KAM_3, FPUT_KAM_4}, Birkhoff-Gustavson normal form analysis \cite{Birkhoff, Gustav, KAM2, Murdock, Wiggins}, and resonance-overlap criteria \cite{FPUT_SST, FPUT_NLin_Res, NLin_Res_Rev, Lichtenberg1, Lichtenberg2, FPUT_Breathers_Res_1, FPUT_Breathers_Res_2} -- together with extensive numerical investigations enabled by modern computational resources.

In the context of the existing FPUT literature, two aspects deserve particular emphasis. First, the energy density considered in the original FPUT simulations is quite small, so that the nonlinear, integrability-breaking perturbation can be regarded as weak. Second, the commonly studied initial conditions -- where only a small fraction of the low-frequency normal modes of the harmonic chain are excited -- are rather atypical.

Early works by Izrailev and Chirikov pointed out the existence of a \textit{stochasticity threshold} in energy density \cite{FPUT_SST}. When this threshold is crossed, nonlinear resonances begin to overlap, giving rise to strong dynamical chaos that rapidly drives the system toward equipartition and thermal equilibrium \cite{FPUT_NLin_Res, Lichtenberg1, Lichtenberg2, FPUT_Breathers_Res_1}. For sufficiently small energy densities, however, the application of KAM theory to FPUT-type systems demonstrates the persistence of regular motion and the absence of equipartition below certain thresholds \cite{FPUT_KAM_1, FPUT_KAM_2, FPUT_KAM_3, FPUT_KAM_4}. A complementary line of investigation, initiated by Zabusky and Kruskal, relates the recurrence phenomenon observed in the FPUT chain to the emergence of solitons in the Korteweg-de Vries (KdV) equation and its truncations in mode space \cite{FPUT_Solitons_1, FPUT_Solitons_2, FPUT_Solitons_3, FPUT_Solitons_4, FPUT_Solitons_5}.   

Furthermore, the commonly used initial conditions often lie close to the so-called $q$-breather solutions, which are exact time-periodic stationary solutions of the FPUT chain \cite{FPUT_Breathers_1, FPUT_Breathers_2, Breathers_Rev_1, Breathers_Rev_2}. This proximity to near-integrable regions of phase space (see also \cite{FPUT_Poggi}) gives rise to two distinct dynamical timescales, a phenomenon often referred to as \textit{metastability} or \textit{prethermalization} \cite{FPUT_Pretherm_Italian_1, FPUT_Pretherm_Italian_2, FPUT_Pretherm_Italian_3, FPUT_Pretherm_Italian_4, FPUT_Pretherm_Italian_5, FPUT_Pretherm_Flach_1, FPUT_Pretherm_Flach_2}. On the shorter timescale, the evolution remains effectively nonergodic, with energy confined to a small set of modes. Only on a much longer timescale does the system gradually escape this near-integrable region and recover the normal statistical behavior associated with thermal equilibrium.

In this paper, we aim to relate the non-KAM dynamics (the relevance of the non-KAM regime is explained in more detail below) observed in the FPUT chain and related anharmonic lattices to nearby integrable dynamics. To remain close to the original FPUT setting, we focus on the FPUT-$\alpha$ model, which contains only the cubic nonlinearity. Besides the trivial integrable limit provided by the linear chain, another closely related integrable system is the Toda chain \cite{Toda1, Toda2, Flaschka1, Manakov, Flaschka2, Henon, Cyprus, Shastry}. Several studies have compared the approach to thermal equilibrium in the initially metastable or prethermal regime of the FPUT-$\alpha$ dynamics with the evolution observed in both the linear and the Toda chains. In particular, the similarity between the metastable FPUT dynamics and the integrable evolution of the Toda chain was first pointed out in Ref.~\cite{FPUT_Int_Dyn_01}. Subsequent works have carried out increasingly detailed comparisons between these models \cite{FPUT_Int_Dyn_02, FPUT_Int_Dyn_03, FPUT_Int_Dyn_04, FPUT_Int_Dyn_05, FPUT_Int_Dyn_1, FPUT_Int_Dyn_2, FPUT_Int_Dyn_3, FPUT_Int_Dyn_4}.

All of these works \cite{FPUT_Int_Dyn_01, FPUT_Int_Dyn_02, FPUT_Int_Dyn_03, FPUT_Int_Dyn_04, FPUT_Int_Dyn_05, FPUT_Int_Dyn_1, FPUT_Int_Dyn_2, FPUT_Int_Dyn_3, FPUT_Int_Dyn_4} largely follow the traditional statistical-mechanics approach \cite{Khinchin} of monitoring the time evolution of suitable observables, such as the average energy of the normal modes, the normalized spectral entropy, or the total energy. From the behavior of these quantities one then extracts characteristic time scales, such as the ergodization time. It is well known, however, that such analyses can be highly sensitive to the choice of observable and to the specific initial conditions \cite{Baldovin}.

In this work, instead, we characterize the weakly nonintegrable dynamics of the FPUT-$\alpha$ chain and related models using universal indicators of chaos: the maximum Lyapunov characteristic exponent (mLCE), the full Lyapunov spectrum (LS), and the rescaled Kolmogorov–Sinai (KS) entropy \cite{Haenggi, Oseledets, Benettin, Skokos1, CSLS, Geist, PP_Book}. According to Pesin's theorem, the KS entropy can be obtained by summing over all positive Lyapunov exponents \cite{Pesin}. These quantities are particularly attractive because, in addition to being coordinate-independent, they remain invariant under a broad class of transformations \cite{Haenggi}.

We note that Ref.~\cite{FPUT_Int_Dyn_4} already employed the mLCE to analyze various FPUT dynamics and to quantify their proximity to either the linear or the Toda chain. In this sense, our work may be viewed as a complementary investigation and a natural extension of Ref.~\cite{FPUT_Int_Dyn_4}.

More importantly, by examining the scaling behavior of the Lyapunov spectrum and the rescaled KS entropy as the system approaches a given integrable limit, we aim to relate these weakly nonintegrable dynamics to the recently identified universality classes governing the slowing down of thermalization \cite{ClassHeisen, Mithun1, CarloGlass, Mithun2, Lubini, Mithun3, MerabPRL, MerabChaos, Gabriel, Weihua, Xiaodong, Patra}. These classes are broadly categorized as (i) long-range network (LRN) and (ii) short-range network (SRN). In the LRN case, weak breaking of integrability induces long-range couplings between the conserved actions of the corresponding integrable system, whereas in the SRN case the resulting couplings remain local.

It should be noted that the theoretical framework underlying these universality classes assumes a sufficiently large phase space dimension and whatever small but finite energy density so that the KAM theorem no longer constrains the dynamics. In this regime, most invariant tori are destroyed and the motion is expected to occur within a single connected chaotic region of phase space. As a result, typical trajectories explore the same accessible region of phase space, leading to dynamical quantities that are largely independent of the specific initial condition \cite{Mechanics_Book_1, Mechanics_Book_2}.

In our numerical calculations, the accessible phase-space dimension is limited by the computational cost of computing the full LS. Nevertheless, we choose an energy density that, while relatively small, exceeds the stochasticity threshold corresponding to the system sizes considered, see Ref.~\cite{FPUT_Int_Dyn_03}. Operating in this regime, the dynamics is expected to be dominated by a single chaotic component of phase space, so that the LS becomes independent of both the initial condition and the particular fiducial trajectory used in the numerical integration.

Finally, we examine how boundary conditions affect the integrability of the Toda chain. Since integrability is a highly special property, integrable Hamiltonians are typically fine-tuned. For example, the Toda chain with either periodic or fixed boundary conditions is integrable when the nearest-neighbor interaction potential retains the exponential form originally introduced by Toda. This integrability can be established through the Lax pair construction \cite{Flaschka1, Manakov, Cyprus, Int_Toda_BC_1, Int_Toda_BC_2, Int_Toda_BC_3, Int_Toda_BC_4}. In particular, the integrability of the fixed-boundary Toda chain can be understood by embedding it into one half of a periodic Toda chain with antisymmetric initial conditions possessing at least three nodes—two at the boundaries and one at the center \cite{Henon}.

Altering either the boundary conditions or the interaction potential can have dramatic consequences for integrability. In particular, the Toda chain with open boundary conditions is not integrable (see also \cite{TodaOBC_He}), even though the difference between the open and fixed-boundary cases amounts only to two interaction terms at the edges. To elucidate this sensitivity, we consider a harmonic chain with a single FPUT-$\alpha$ nonlinearity introduced at one lattice site. As argued in the appendix, even a single defect can generate a LRN in the space of conserved actions. By interpolating between the open and fixed-boundary Toda chains and analyzing the scaling properties of the LS and the rescaled KS entropy, we demonstrate that thermalization slows down through the LRN mechanism.

The paper is organized as follows. In Sec.~\ref{Sec:Models_Methods}, we introduce the model Hamiltonians, discuss their integrability properties, and describe how the relevant integrable limits are approached. In that section, we also define the LS and the rescaled KS entropy, and outline the numerical integration scheme used in our calculations. Section~\ref{Sec:Chaos_Indicators} presents our numerical results, including a comparison of the dynamics of a harmonic chain with a single FPUT-$\alpha$ defect and an analysis of the LRN that emerges when approaching the fixed-boundary Toda chain from the open-boundary case. Finally, we conclude with a discussion of our findings.

\section{Models and Methods}
\label{Sec:Models_Methods}

In Sec.~\ref{Sec:Models}, we present the model Hamiltonians studied in this work, highlighting the integrable ones and their key properties. Section~\ref{Sec:Methods} introduces the main indicators of chaos, namely the Lyapunov spectrum (LS) and the rescaled Kolmogorov–Sinai (KS) entropy.

\subsection{Model Hamiltonians and Boundary Conditions}
\label{Sec:Models}

We consider 1D nonlinear lattice chains with the following nearest-neighbor interaction potentials: 
\begin{subequations}
\begin{align}
    &V_\textrm{lin}(r) = \frac{1}{2} r^2, \label{V_lin} \\
    &V_\textrm{FPUT-$\alpha$}(r) = \frac{1}{2} r^2 + \frac{\alpha}{3}r^3, \label{V_FPUT} \\
    &V_\textrm{Toda}(r) = \frac{1}{4\alpha^2} \left( e^{2\alpha r} - 2\alpha r - 1 \right), \label{V_Toda} \\ &V_\textrm{exp}(r) = \frac{1}{4\alpha^2} \left( e^{2\alpha r} - 1 \right). \label{V_exp}
\end{align}
\label{NN_Int_Pot}
\end{subequations}
The first interaction potential \eqref{V_lin} corresponds to the harmonic spring potential obeying Hooke’s law. The second potential \eqref{V_FPUT} was introduced by Fermi, Pasta, Ulam, and Tsingou in their celebrated numerical experiment \cite{FPUT}. The third potential was proposed by Toda \cite{Toda1, Toda2}. The last potential corresponds to the bare exponential Toda interaction before the subtraction of the linear term. Writing the nearest interaction potentials as $V_\textrm{int}(r)$, we define the potential of the entire lattice satisfying either the open or fixed-boundary condition as 
\begin{subequations}
\begin{align}
    &V_\textrm{lattice, FBC} = V_\textrm{int}\left(q_{1}\right) + \left[\sum_{i=1}^{N-1} V_\textrm{int}\left(q_{i+1} - q_{i}\right)\right]  \label{FBC_Chain} + V_\textrm{int}\left(- q_{N}\right), \\
    &V_\textrm{lattice, PBC} =  \left[\sum_{i=1}^{N} V_\textrm{int}\left(q_{i+1} - q_{i}\right)\right] \textrm{ with $q_{i+N} = q_i$,}  \label{PBC_Chain} \\
    &V_\textrm{lattice, OBC} = \sum_{i=1}^{N-1} V_\textrm{int}\left(q_{i+1} - q_{i}\right),
    \label{OBC_Chain}
\end{align}
\label{Latt_Pot}
\end{subequations}
where $q_i$ denotes the displacement of the $i$th mass from equilibrium. This implies that 
\begin{equation}
    V_\textrm{lattice, FBC} = V_\textrm{lattice, OBC} + \left[ V_\textrm{int}\left(q_{1}\right) + V_\textrm{int}\left(- q_{N}\right) \right]. 
\label{FBC_OBC}
\end{equation}
We write 1D lattice Hamiltonians using the lattice potential \eqref{Latt_Pot} introduced above as
\begin{equation}
\begin{split}
    H_\textrm{lattice}(\bm{q}, \bm{p}) = \left[\sum_{i = 1}^N \frac{p_i^2}{2}\right] + V_\textrm{lattice}(\bm{q}) 
    \equiv T_\textrm{lattice}(\bm{p}) + V_\textrm{lattice}(\bm{q}),
\end{split}
    \label{Latt_Ham}
\end{equation}
where the canonical coordinates and momenta for the $N$ degrees of freedom are denoted by $\bm{q} = \{q_1, \ldots, q_N\}$ and $\bm{p} = \{p_1, \ldots, p_N\},$ with $p_i$ representing the momentum conjugate to the lattice displacement $q_i$. Here $T_\textrm{lattice}(\bm{p})$ denotes the kinetic energy of the lattice.

The integrability of the harmonic chain with interaction potential \eqref{V_lin}, under fixed, periodic, or open-boundary conditions, is most directly established by constructing the action–angle variables, which are obtained through the normal mode decomposition (see Appendix~\ref{App_SHO_1FPUT}). The FPUT-$\alpha$ chain, in contrast, is not integrable. 

When Toda first introduced his eponymous lattice, he obtained a number of exact soliton solutions \cite{Toda1, Toda2}. The complete integrability of the periodic Toda lattice was later established through the construction of a Lax representation \cite{Flaschka1, Manakov, Lax}. To derive this representation, one begins with the $2N$ Hamilton equations of motion
\begin{equation}
    \dot{{\bm q}}= \frac{\partial H}{\partial {\bm p}}\ ,\quad
    \dot{{\bm p}}=- \frac{\partial H}{\partial {\bm q}},
    \label{Ham_Dyn_Eqns}
\end{equation}
which can be compactly written as
\begin{equation}
    \dot{{\bm z}} = L_H{\bm z} = \{H,{\bm z}\}, 
    \label{Liouvillian}
\end{equation}
where $\bm{z} = \left(\bm{q}, \bm{p}\right)$, and the Liouvillian operator $L_H$ is defined through the Poisson bracket
\begin{equation}
    \label{Poisson}
    \{A,B\}= \frac{\partial A}{\partial {\bm q}} \frac{\partial B}{\partial {\bm p}} - \frac{\partial B}{\partial {\bm q}} \frac{\partial A}{\partial {\bm p}}.
\end{equation}
Introducing the Flaschka variables \cite{Flaschka1}
\begin{equation}
    \label{Flaschka_Var}
    a_j = \frac{1}{2}e^{\alpha\left(q_{j+1} - q_j\right)}, \quad b_j = \alpha p_j,
\end{equation}
the equations of motion derived from Eq.~\eqref{Ham_Dyn_Eqns} take the form
\begin{equation}
    \dot{a}_j = a_j\left(b_{j+1} - b_{j}\right), \quad \dot{b}_{j} = 2\left(a_{j}^2 - a_{j-1}^2\right).
    \label{Flaschka_Dyn_Eqns}
\end{equation}
These equations can be recast as the Lax equation \cite{Flaschka1, Manakov, Lax}
\begin{equation}
    \dot{L} = \left[B, L\right] \equiv BL - LB,
    \label{Lax_Pairs}
\end{equation}
where $L(\bm{a}, \bm{b})$ is a symmetric $N \times N$ matrix and $B(\bm{a}, \bm{b})$ is a skew-symmetric matrix function of the same dimension. 

It follows that $L(t)$ remains similar to $L(0)$, implying that the eigenvalues of $L$ are conserved in time. It is therefore convenient to introduce the conserved quantities
\begin{equation}
    I_{k} = \frac{1}{k!}\mathrm{Trace} \left(L^{k}\right), \qquad 1 \le k \le N.
    \label{Toda_Int_Lax}
\end{equation}
which form a complete set of integrals of motion. One readily verifies that $I_1$ corresponds to the total momentum, while $I_2$ equals the total energy up to an additive constant for the periodic Toda chain. The higher integrals become increasingly complicated and correspond to nonlocal quantities in the original lattice variables. 

In fact, a set of action-angle variables was constructed in Ref.~\cite{Flaschka2} using the \textit{discriminant} of the Lax matrix $L$, denoted by $\Delta(\lambda)$, defined through
\begin{equation}
\mathrm{det}\left(\lambda\mathbb{1} - L\right) = \left[\prod_{j=1}^{N} a_j \right]\left[\Delta(\lambda) - 2\right].
\label{Discriminant_Lax}
\end{equation}
The $k$th action variable can then be written (see also \cite{FPUT_Int_Dyn_01}) as 
\begin{equation}
J_{k} = \frac{2}{\pi} \int_{\Lambda_{2k+1}}^{\Lambda_{2k}} \cosh^{-1}\left(\frac{\left|\Delta(\lambda)\right|}{2}\right) d\lambda,
\label{Toda_Actions}
\end{equation}
where $\Lambda_{k}$ are the consecutively ordered roots of $\Delta(\lambda) = \pm 2$. These action variables are manifestly nonlocal when expressed in terms of the original lattice variables.

Following Ref.~\cite{Henon}, the integrals of motion for the fixed-boundary Toda chain with $N$ lattice sites can be obtained by embedding it into a periodic lattice with $(2N + 2)$ sites labeled by $j = -N, \ldots, N+1$. On this extended periodic lattice we impose the antisymmetric initial conditions
\begin{equation}
\begin{split}
q_{-j}=-q_j,\quad p_{-j}=-p_j, \quad q_0=q_{N+1}=0,\quad p_0=p_{N+1}=0,
\end{split}
\label{Antisymm_Init}
\end{equation}
for $j = 1, \ldots, N,$ where $q_j$ and $p_j$ are otherwise arbitrary for these indices. Because the equations of motion preserve this antisymmetry, the segment of the periodic lattice between $j = 1$ and $j = N$ evolves exactly as a fixed-boundary Toda chain with $N$ sites. The periodic system possesses $2(N+1)$ integrals of motion, which we denote by $I_{k}^\prime$ for $k = 1, 2, \ldots, 2(N+1)$, where the prime indicates that the antisymmetric constraint \eqref{Antisymm_Init} has been imposed. Under this constraint all integrals with odd $k$ vanish identically, while one additional integral reduces to a constant independent of the dynamical variables. Consequently, the fixed-boundary Toda chain is integrable and is characterized by the $N$ nontrivial integrals $I^\prime_{k}$ with $k = 2, 4, \ldots, 2N$.  

Having discussed the integrability of the periodic and fixed-boundary Toda chains, we now turn to the open-boundary case. Before proceeding, we note the following relations between the Toda and bare exponential interaction potentials:
\begin{subequations}
\begin{align}
    V_\textrm{Toda, PBC} &= V_\textrm{exp, PBC}, \\
    V_\textrm{Toda, FBC} &= V_\textrm{exp, FBC}, \\
    V_\textrm{Toda, OBC} &= V_\textrm{exp, OBC} + \frac{1}{2\alpha}\left(q_1 - q_N\right). \label{Toda_Exp_Relation_OBC}
\end{align}
\label{Toda_Exp_Relation}
\end{subequations}
It is well known that the bare exponential Toda lattice is integrable under all three boundary conditions. For open boundaries the Lax matrices $L$ and $B$ are tridiagonal \cite{Cyprus}. In the periodic case they remain tridiagonal in the bulk but acquire additional corner elements $L_{1N} = L_{N1} = -B_{1N} = B_{N1} = a_{N}$ \cite{Flaschka1}. The same Lax construction fails for $H_\textrm{Toda, OBC}$ because of the additional linear term in Eq.~\eqref{Toda_Exp_Relation_OBC}; see \cite{TodaOBC_He} for a simple proof. Consequently, the open-boundary Toda lattice is not integrable, a conclusion also supported by the presence of a positive maximum Lyapunov characteristic exponent. Notably, unlike the FPUT-$\alpha$ model -- where where all intersite interactions differ from those of the integrable Toda chain (e.g., the fixed-boundary one) -- the open-boundary Toda lattice departs from the integrable case only through two boundary interaction terms, see Eq.~\eqref{FBC_OBC}.

\subsection{Lyapunov Spectrum and Kolmogorov–Sinai Entropy}
\label{Sec:Methods}

We quantify the nonintegrability -- manifested as the sensitive dependence of trajectories on initial conditions -- by computing the Lyapunov spectrum (LS) and related chaotic indicators for the Hamiltonians considered here. As noted in the introduction, these quantities possess universal properties: their asymptotic values are invariant under smooth canonical transformations and are largely independent of the choice of initial conditions \cite{Haenggi}. In this section, we briefly review the relevant theoretical results and describe the procedure used to compute them \cite{CSLS, PP_Book}.

To compute the Lyapunov characteristic exponents (LCEs), we first derive the evolution equation for infinitesimal perturbations by linearizing Hamilton’s equations \eqref{Liouvillian}. This yields the variational equation 
\begin{equation}
\label{eq:var_eq}
    \dot{{\bm w}}(t) =\left[ \mathbb{\Omega}_{2N} \cdot \mathbb{D}_{H}^2 ({\bm z}(t) ) \right] \cdot {\bm w} (t),
\end{equation}
where  ${\bm w}(t) = (\delta {\bm q}(t), \delta {\bm p}(t))=\left(\delta q_1 (t), \ldots, \delta q_N (t), \delta p_1(t), \ldots, \delta p_N (t) \right)$ is an infinitesimal deviation from the trajectory ${\bm z}(t) = \left( {\bm q}(t), {\bm p}(t) \right)$. The matrix $\mathbb{D}_{ H}^2 ({\bm z}(t) )$ denotes the Hessian of the Hamiltonian evaluated along the trajectory, 
\begin{equation}
    \label{eq:hessian}
    \left[ \mathbb{D}_{H}^2 ({\bm z}(t) ) \right]_{i, j} =
    \frac{\partial^2 H}{\partial z_i \partial z_j}\bigg\vert_{{\bf z}(t)},\qquad i,j=1,\ldots,2N,
\end{equation}
while the symplectic identity matrix is
\begin{equation}
\label{eq:symp_max}
\mathbb{\Omega}_{2N} =
\begin{bmatrix}
    \mathbb{O}_N  &  \mathbb{1}_N \\
   - \mathbb{1}_N  &  \mathbb{O}_N
\end{bmatrix},
\end{equation}
where $\mathbb{1}_N$ and $\mathbb{O}_N$ denote $N \times N$ identity and null matrices, and  $\mathbb{\Omega}^2_{2N} = -\mathbb{1}_{2N}$. From Eq.~\eqref{eq:var_eq}, the generator of the tangent dynamics is
\begin{equation}
\label{Gen_Jacobian}
\mathbb{A}({\bm z}, t) \equiv \mathbb{\Omega}_{2N} \cdot\mathbb{D}{H}^2 ({\bm z}(t)).
\end{equation}
Since the Hessian is symmetric, $\mathbb{D}_{H}^2 ({\bm z}(t) ) = \left[\mathbb{D}_{H}^2 ({\bm z}(t) )\right]^\mathrm{T}$, it follows that
\begin{equation}
    \label{Sympl_Lie}
    \mathbb{A}^\mathrm{T}({\bm z}, t)\cdot\mathbb{\Omega}_{2N} + \mathbb{\Omega}_{2N} \cdot \mathbb{A}({\bm z}, t)   = 0, 
\end{equation}
where we used $\mathbb{\Omega}^2_{2N} =  - \mathbb{1}_{2N}$. Hence $\mathbb{A}({\bm z}, t)$ belongs to the symplectic Lie algebra $\mathfrak{sp}(2N,\mathbb{R})$. For a short time step $\delta t$, the infinitesimal Jacobian
\begin{equation}
    \label{Inf_Jac}
    \mathbb{J}({\bm z}, t) \approx \exp\left[ \mathbb{A}({\bm z}, t) \delta t \right],
\end{equation}
is therefore symplectic, i.e., $\mathbb{J}({\bm z}, t) \in Sp(2N, \mathbb{R})$, and satisfies 
\begin{equation}
    \label{Jac_Sympl}
    \mathbb{J}({\bm z}, t) \cdot \mathbb{\Omega}_{2N} \cdot \mathbb{J}^\mathrm{T}({\bm z}, t) = \mathbb{\Omega}_{2N}.
\end{equation}

The variational equations \eqref{eq:var_eq} form a system of linear differential equations for ${\bm w}$. Their main complication lies in the time dependence of the coefficients through the instantaneous generator $\mathbb{A}({\bm z}, t)$. The solution at $t = T_\mathrm{end}$ can nevertheless be obtained by integrating the equations, yielding 
\begin{equation}
\label{Sol_Var_Eqns}
    \begin{split}
        {\bm w} (T_\mathrm{end}) = \mathbb{J}({\bm z}_0, T_\mathrm{end}) \cdot {\bm w}(0),  
    \end{split}
\end{equation}
where ${\bm z}(0) \equiv {\bm z}_0$ is the initial phase-space point and the Jacobian matrix $\mathbb{J}({\bm z}_0, T_\mathrm{end})$ is expressed through the time-ordered exponential $\mathcal{T}\exp$ as in   
\begin{equation}
\label{Int_Jac}
    \begin{split}
        \mathbb{J}({\bm z}_0, T_\mathrm{end}) = \mathcal{T}\exp\left[\int_{0}^{T_\mathrm{end}}dt\; \mathbb{J}({\bm z}(t), t)\right].  
    \end{split}
\end{equation}
For the numerical computation of the LCEs, the full Jacobian is approximated as in 
\begin{equation}
\label{Full_Jac_Apprx}
    \begin{split}
        \mathbb{J}({\bm z}_0, T_\mathrm{end}) = \lim_{N_\mathrm{step} \to \infty}\prod_{k=0}^{N_\mathrm{step} - 1} \mathbb{J}\left({\bm z}(k\delta t), k\delta t\right) 
        = \lim_{N_\mathrm{step} \to \infty}\prod_{k=0}^{N_\mathrm{step} - 1}\exp\left[ \mathbb{A}\left({\bm z}(k\delta t), k\delta t\right) \delta t \right],  
    \end{split}
\end{equation}
with $T_\mathrm{end} = N_\mathrm{step} \delta t$ and $\delta t \to 0$ the integration time step. Using Eq.~\eqref{Full_Jac_Apprx} together with the fact that the product of symplectic matrices is symplectic, it follows that the full Jacobian $\mathbb{J}({\bm z}_0, T_\mathrm{end})$ of the tangent flow is also symplectic.  

Since the LCEs characterize the asymptotic average rate of growth (or decay) of infinitesimal perturbations, we consider the norm squared of the perturbation vectors ${\bm w}(t)$. From 
\begin{equation}
\label{w_Norm}
    \lVert {\bm w}(t) \rVert = {\bm w}^\mathrm{T}(0)  \cdot \mathbb{J}^\mathrm{T}({\bm z}_0, T_\mathrm{end})  \cdot \mathbb{J}({\bm z}_0, T_\mathrm{end}) \cdot {\bm w}(0),  
\end{equation}
the evolution of ${\bm w}(t)$ is governed by the real, symmetric, positive-definite matrix defined in 
\begin{equation}
\label{M_Matt_Defn}
    \begin{split}
        \mathbb{M}({\bm z}_0, T_\mathrm{end}) = \mathbb{J}^\mathrm{T}({\bm z}_0, T_\mathrm{end}) \cdot \mathbb{J}({\bm z}_0, T_\mathrm{end}). 
    \end{split}
\end{equation}
The Oseledets Multiplicative Ergodic Theorem (MET) \cite{Oseledets} then states that, for a statistically stationary and ergodic sequence of matrices generated by the underlying ergodic flow ${\bm z}(t)$, the limit in
\begin{equation}
\label{P_Defn_MET}
    \begin{split}
        \lim_{T_\mathrm{end}\to \infty} \left[\mathbb{M}({\bm z}_0, T_\mathrm{end})\right]^{1/2T_\mathrm{end}} = \mathbb{P} 
    \end{split}
\end{equation}
exists. In the non-KAM dynamical regime of interest, where the chaotic region of phase space is fully connected, the matrix $\mathbb{P}$ becomes independent of the initial phase-space point. It has $2N$ positive eigenvalues $\mu_1 \geqslant  \mu_2 \geqslant \ldots \geqslant \mu_{2N}$, from which the Lyapunov spectrum is obtained via    
\begin{equation}
    \Lambda_{k} = \ln \mu_k.
    \label{Lya_Spect_MET}
\end{equation}
for $k = 1, \ldots, 2N$. 

\begin{figure}
    \centering
    \includegraphics[trim={0.75cm 0.55cm 0cm 0cm},clip,scale=0.85]{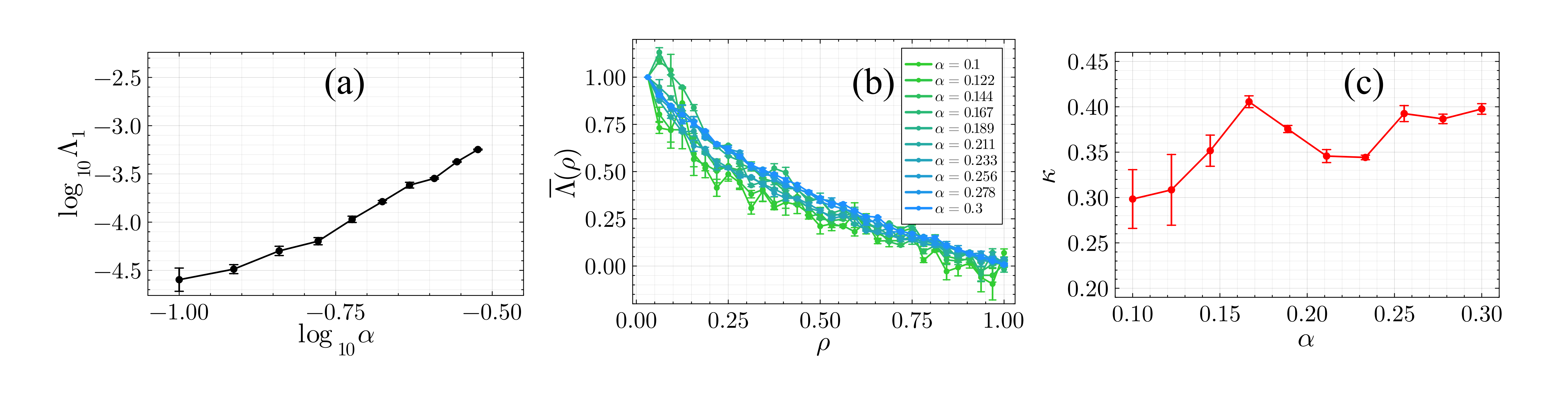}
    \caption{ Fixed boundary condition FPUT-$\alpha$ with $N = 32$. We perform the calculations starting from an initial condition with all lattice displacements set to zero, while the lattice momenta are randomly drawn subject to the constraint that the energy density equals 0.1. As the integrable limit $H_\textrm{lin,FBC}$ is approached by decreasing $\alpha$, the maximum Lyapunov characteristic exponent (mLCE) $\Lambda_1$ steadily decreases, as shown in panel (a). In panel (b), we plot the rescaled Lyapunov spectra (LS) $\overbar{\Lambda}(\rho) \equiv \Lambda_i/\Lambda_1$ as a function of $\rho = i/N$, with $i = 1, \ldots, N$, for values of $\alpha$ equally spaced between 0.1 and 0.3. The area under these spectra appears to saturate to a finite value, consistent with the behavior of the rescaled Kolmogorov–Sinai entropy $\kappa$ shown in panel (c). This scaling of the Lyapunov spectra indicates that thermalization slows down according to the long-range network (LRN) scenario.}
    \label{Fig:FPUT_FBC}
\end{figure}

\begin{figure}
    \centering
    \includegraphics[trim={0.75cm 0.55cm 0cm 0cm},clip,scale=0.85]{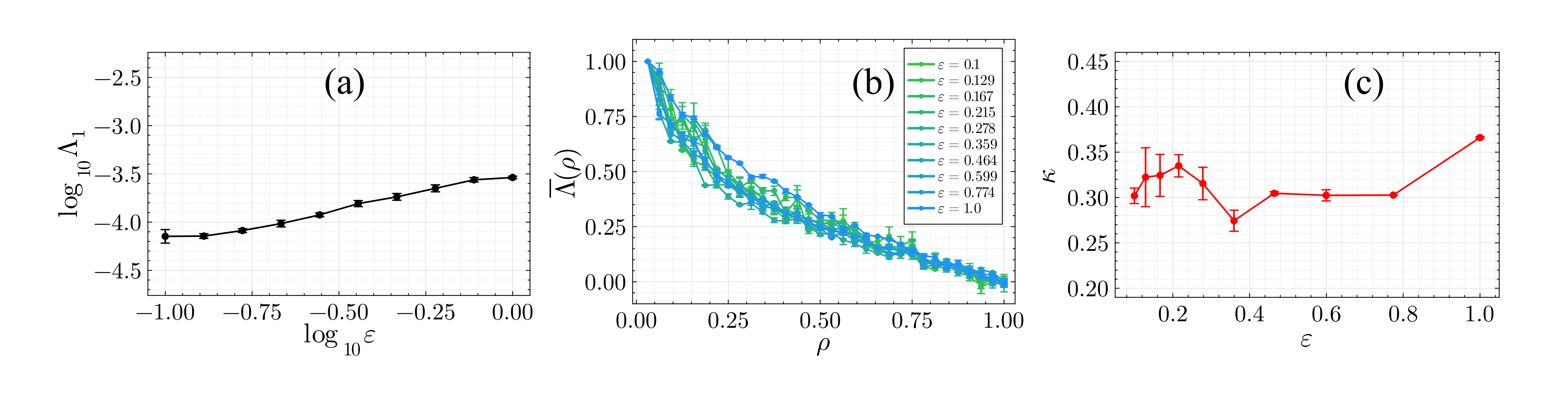}
    \caption{ We interpolate between the nonintegrable FPUT-$\alpha$ chain and the integrable Toda chain by tuning the parameter $\varepsilon$, imposing fixed boundary conditions in both cases. The potential is given by Eq.\ \eqref{V_Mixed_FBC} with $N = 32$, and the initial condition is the same as in Fig.~\ref{Fig:FPUT_FBC}. The parameter $\varepsilon$ is sampled at logarithmically spaced values between 0.1 and 1.0. As in Fig.~\ref{Fig:FPUT_FBC}, panels (a), (b), and (c) show the maximal Lyapunov characteristic exponent (mLCE) $\Lambda_1$, the rescaled Lyapunov spectrum $\overbar{\Lambda}(\rho) \equiv \Lambda_i/\Lambda_1$ as a function of $\rho = i/N$ $(i = 1, \ldots, N)$, and the rescaled Kolmogorov–Sinai (KS) entropy $\kappa$, respectively. As before, the results indicate that thermalization slows down in accordance with the long-range network (LRN) scenario. }
    \label{Fig:Mixed_FBC}
\end{figure}

\begin{figure}
    \centering
    \includegraphics[trim={0.75cm 0.55cm 0cm 0cm},clip,scale=0.85]{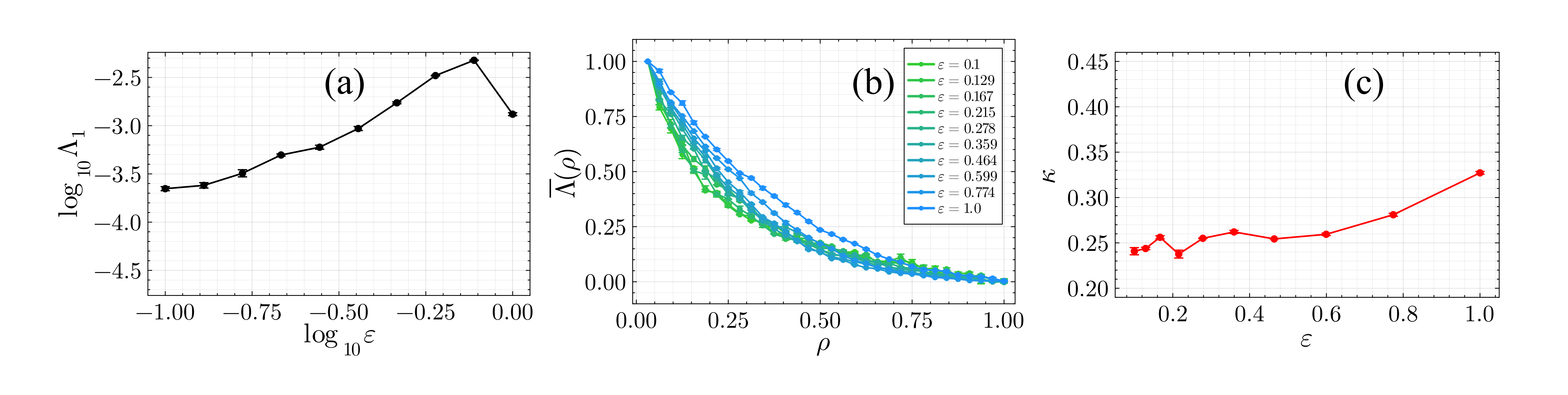}
    \caption{ Mixed boundary condition Toda chain with $N = 32$. The potential is taken as in Eq.~\eqref{V_Toda_MBC}, whose integrable limit corresponds to the fixed-boundary Toda chain. The initial condition is identical to that used in Fig.~\ref{Fig:FPUT_FBC}. The parameter $\varepsilon$ is varied over logarithmically spaced values in the interval $0.1 \leqslant \varepsilon \leqslant 1.0$. Panels (a), (b), and (c) display the maximal Lyapunov characteristic exponent (mLCE) $\Lambda_1$, the rescaled Lyapunov spectrum $\overbar{\Lambda}(\rho) \equiv \Lambda_i/\Lambda_1$ as a function of $\rho = i/N$ $(i = 1, \ldots, N)$, and the rescaled Kolmogorov–Sinai (KS) entropy $\kappa$, respectively. The results again support the picture that thermalization becomes progressively slower, consistent with the long-range network (LRN) mechanism. }
    \label{Fig:Toda_MBC}
\end{figure}

Because the Jacobian is symplectic, if $\mu$ is an eigenvalue of $\mathbb{M}$, then so is $1/\mu$. Consequently, each positive Lyapunov exponent \( \Lambda_i \) is paired with a negative one, \( \Lambda_{2N - i +1} = - \Lambda_i \), reflecting the preservation of phase-space volume. It is therefore sufficient to analyze the rescaled positive LCEs \(\overbar{\Lambda}_{i} \equiv \Lambda_i/\Lambda_1\) to identify the universality classes associated with thermalization slowdown \cite{MerabPRL, Gabriel}. Accordingly, we plot the normalized spectrum \( \overline{\Lambda}(\rho) \) versus \( \rho = i/N \) in Figs.~\ref{Fig:FPUT_FBC}, \ref{Fig:Mixed_FBC}, and \ref{Fig:Toda_MBC}. Moreover, the key features relevant for understanding thermalization slowdown are succinctly captured by the behavior of the rescaled Kolmogorov-Sinai (KS) entropy,
\begin{equation}
    \label{KS_Entropy}
    \kappa = \frac{1}{N}\sum_{i = 1}^{N} \overbar{\Lambda}_{i}  \xrightarrow[N\to\infty]{} \int_{0}^{1} \overbar{\Lambda}(\rho)\, d\rho, 
\end{equation}
obtained from the LS.

We conclude this section by summarizing the expected behavior of the LCEs in the two universality classes of thermalization slowdown \cite{Gabriel}. In the long-range network (LRN) class, the maximal exponent $\Lambda_1$ vanishes -- typically as a power law -- while the rescaled spectrum $\overline{\Lambda}(\rho)$ converges to a limiting curve that itself decays as a power law in $\rho$ as the integrable limit is approached. Consequently, the rescaled KS entropy $\kappa$ approaches a finite constant. 

In contrast, in the short-range network (SRN) class, although $\Lambda_1$ again vanishes algebraically, the rescaled spectra do not converge. Instead, they are well described by an exponential form $\overline{\Lambda}(\rho) \sim e^{-\beta \rho}$, with $\beta$ diverging as the integrability-breaking parameter tends to zero. Correspondingly, $\kappa \to 0$ as $1/\beta$ in the integrable limit.

\subsection{Computational Framework: Split-Step Integration and Lyapunov Analysis}
\label{Sec:Methods_Algorithm}

In our numerical algorithm for computing the LS \cite{Benettin, Skokos1, CSLS, Geist, PP_Book}, we first consider the finite-time maximal Lyapunov characteristic exponent of order \( p \), denoted by \( \Lambda^{p}(t) \). This quantity measures the exponential rate at which the volume of a \( p \)-dimensional parallelogram, spanned by \( p \) linearly independent deviation vectors \( \bm{w}_1(t), \bm{w}_2(t), \ldots, \bm{w}_p(t) \), evolves in time. It is defined as in
\begin{equation}
\Lambda^{p}(T_\mathrm{end}) = \frac{1}{T_\mathrm{end}} \sum_{j = 1}^{N_\mathrm{step}} \ln\left(\frac{\text{vol}_p[\bm{w}_1(j\delta t), \bm{w}_2(j\delta t), \ldots, \bm{w}_p(j\delta t)]}{\text{vol}_p[\bm{w}_1((j-1)\delta t), \bm{w}_2((j-1)\delta t), \ldots, \bm{w}_p((j-1)\delta t)]}\right), \label{ft_pmLCE}
\end{equation}

where \( \text{vol}_p[\cdot] \) denotes the volume of the \( p \)-parallelogram formed by the given vectors. Taking the infinite-time limit,
\begin{equation}
    \Lambda_1^{p} = \lim_{T_\mathrm{end} \to \infty} \Lambda^{p}(T_\mathrm{end}),
    \label{p-mLCE}
\end{equation}
yields the maximal Lyapunov characteristic exponent of order $p$ ($p$-mLCE).

Starting from an orthonormal set of deviation vectors $\{\bm{w}_1(0), \bm{w}_2(0), \ldots, \bm{w}_p(0)\}$, their time evolution over a step $\delta t$ produces the set $\{\bm{w}_1(\delta t), \bm{w}_2(\delta t), \ldots, \bm{w}_p(\delta t)\}$, which is generally no longer orthonormal. The vectors must therefore be reorthonormalized before further evolution. This can be achieved through Gram–Schmidt orthonormalization and applied iteratively while following a fiduciary trajectory ${\bm z}(t)$. Besides preventing numerical overflow, this renormalization aligns the deviation vectors with the most unstable directions, allowing the $p$-mLCEs to converge more rapidly. 

In practice, however, the Gram–Schmidt procedure may suffer from numerical instabilities. Therefore, following Refs.~\cite{CSLS, Geist, PP_Book, Gabriel}, we instead perform a QR decomposition of the $2N \times p$ matrices defined in 
\begin{equation}
{\bm W}(j\delta t) = \begin{bmatrix}
    {\bm w}_1(j\delta t) & {\bm w}_2(j\delta t) & \cdots & {\bm w}_{p}(j\delta t)
\end{bmatrix}, 
\label{W_Mat_QR}
\end{equation}
at each step $j = 1, 2, \ldots, N_\mathrm{step}$. This provides a numerically stable way to obtain the expansion coefficients of the evolving $p$-parallelograms along the fiducial trajectory, without explicitly computing their spanning vectors.

The individual LCEs, which together form the full LS, are then obtained from
\begin{equation}
    \Lambda_p = \Lambda_1^{p} - \Lambda_1^{p-1},
    \label{LCE_p-mLCE}
\end{equation}
with \(\Lambda_1^{0} \equiv \Lambda_1\) the maximal LCE computed using 
\begin{equation}
    \Lambda_1 = \lim_{N_\mathrm{step} \to \infty}\frac{1}{N_\mathrm{step} \delta t} \sum_{j = 1}^{k} \ln \frac{ \lVert \bm{w}(j\delta t) \rVert }{ \lVert \bm{w}( (j-1)\delta t ) \rVert }. \label{Eq:mLCE}
\end{equation}
Here \( \lVert \bm{w}(j\delta t) \rVert \) denotes the magnitude of the deviation vector at time $t = j\delta t$, where $\delta t$ is the step size used in evaluating all \( \Lambda_1^{p} \).

We employ split-step symplectic integration schemes to compute the fiducial trajectory from Hamilton's equations. In these methods, the Hamiltonian is decomposed into exactly solvable parts, and the time evolution is constructed through successive applications of their individual propagators \cite{Trot, Trot_Sympl_1, Neri, Yoshida, Koseleff1, McLachlan1, Koseleff2, McLachlan2, Laskar1, Laskar2, Tao, Skokos2, AP}. Formally, the solution of Eq.~\eqref{Liouvillian} can be written as in
\begin{equation}
\label{eq:ham_eq_sol}
{\bm z}(t) = e^{t L_H}{\bm z}(0)
= \sum_{s=0}^{+\infty} \frac{t^s}{s!} \left(L_H\right)^s {\bf z}(0).
\end{equation}

In split-step schemes, the operator $e^{t L_H}$ is approximated so that phase-space volume is preserved at each time step; consequently, every step of the map corresponds to a canonical transformation. The lattice Hamiltonians considered here can be written as $H= A + B$, where $A$ is the total kinetic energy and $B$ is the total potential energy. For this decomposition, the operators $e^{tL_A}$ and $e^{tL_B}$ are known explicitly in closed form.

Using the Baker–Campbell–Hausdorff formula, the operator $e^{\delta t L_H}$ is approximated as in 
\begin{equation}
e^{\delta t L_H} = \prod_{j=1}^k  e^{ a_j L_{A} \delta t }e^{b_j L_{B} \delta t }  + \mathcal{O}(\tau^{p}),
\label{eq:symp}
\end{equation}
where the coefficients $a_1,b_1,a_2,b_2,...,a_k,b_k$ satisfy the consistency conditions $\sum_{j=1}^k a_j = \sum_{j=1}^k b_j = 1$. The accuracy of the approximation is determined by the order $p$, which depends on $k$ and and on the choice of these coefficients. 

We compute the LS using the tangent map method of Ref.~\cite{Skokos1}, in which the same split-step integrator is applied to both the phase-space trajectory and the variational flow. The flow equations corresponding to the $A$ and $B$ steps are given in
\begin{equation}
\label{Tangent_Map_Split-Step}
\begin{split}
    \left.
        \begin{aligned}
                &\dot{\bm{q}} = \bm{p}, \\
                &\dot{\bm{p}} = 0, \\
                &\dot{\delta\bm{q}} = \bm{\delta p}, \\
                &\dot{\delta\bm{p}} = 0,
        \end{aligned}
    \right\} \Rightarrow \frac{d\bm{z}}{dt} = L_{A}\bm{z}, \qquad  
    \left.
        \begin{aligned}
                &\dot{\bm{q}} = 0, \\
                &\dot{\bm{p}} = - \frac{ \partial V\left( \bm{q} \right) }{ \partial \bm{q} }, \\
                &\dot{\delta\bm{q}} = 0, \\
                &\dot{\delta\bm{p}} = - \mathbb{D}^{2}_{V}\left( \bm{q} \right) \delta\bm{q},
        \end{aligned}
    \right\} \Rightarrow \frac{d\bm{z}}{dt} = L_{B}\bm{z}, 
\end{split}
\end{equation}
where the Hessian of the potential,
\begin{equation}
\label{Hess_Pot}
    \mathbb{D}^{2}_{V}\left( \bm{q}(t) \right)_{jk} = \frac{ \partial^2 V\left( \bm{q} \right) }{ \partial q_j \partial q_k } \bigg|_{\bm{q}(t)}, \quad j,k = 1, 2, \ldots, N,
\end{equation}
appears.

The $A$ step reduces to a system of simple ordinary differential equations (ODEs) with constant coefficients and can therefore be solved straightforwardly. During the $B$ step, over the interval $\left[ t_0, t_{0} + b_{j}\delta t \right]$, the coordinates $\bm{q}$ remain constant. Consequently, the gradient and Hessian of the potential,
\begin{equation}
\label{Grad_Hess_Pot_Const}
    \frac{ \partial V\left( \bm{q} \right) }{ \partial \bm{q} } =  \frac{ \partial V\left( \bm{q} \right) }{ \partial \bm{q} }\bigg|_{\bm{q}(t) = \bm{q}(t_0)}, \qquad\mathbb{D}^{2}_{V}\left( \bm{q}(t) \right) = \frac{ \partial^2 V\left( \bm{q} \right) }{ \partial q_j \partial q_k } \bigg|_{\bm{q}(t) = \bm{q}(t_0)}, \; j,k = 1, \ldots, N,
\end{equation}
remain constant, so the corresponding flow equations reduce to ODEs with constant coefficients. This considerably simplifies the implementation, since in both steps one only needs to solve linear systems of ODEs with constant coefficients.

The fourth-order symplectic integration scheme used in this work is the $ABA864$ method, first introduced in Ref.~\cite{Laskar2}. Following Ref.~\cite{Laskar2}, we express $e^{\delta t L_H}$ as in 
\begin{equation}
\begin{aligned}
  e^{\delta t L_H} = &e^{a_1 L_{A} \delta t }e^{b_1 L_{B} \delta t }e^{a_2 L_{A} \delta t } e^{b_2 L_{B} \delta t } e^{a_3 L_{A} \delta t }e^{b_3 L_{B} \delta t } \\
  &\times e^{a_4 L_{A} \delta t } e^{b_4 L_{B} \delta t} e^{a_4 L_{A} \delta t } \\ 
  &\times e^{b_3 L_{B} \delta t } e^{a_3 L_{A} \delta t }e^{b_2 L_{B} \delta t } e^{a_2 L_{A} \delta t }e^{b_1 L_{B} \delta t } e^{a_1 L_{A} \delta t }\\
  &+ \mathcal{O}\left(\delta t^4\right) 
\end{aligned}
\label{eq:aba864}
\end{equation}
for a time step $\delta t$. Note the palindromic sequence formed by the splitting coefficients, ensuring symmetry and exact time reversibility. Truncating the values reported in Table 3 of Ref.~\cite{Laskar2} to eight decimal places, the coefficients in Eq.~\eqref{eq:aba864}, $\{a_1,a_2,a_3,a_4,b_1,b_2,b_3,b_4\}$, are given in (see also \cite{Skokos2, AP})  
\begin{equation}
\begin{split}
&a_1 = 0.07113343,  
\qquad \qquad \ \ 
b_1 = 0.18308368, \\ 
&a_2 = 0.24115343,  
\qquad \qquad \ \ 
b_2 = 0.31078286, \\ 
&a_3 = 0.52141176,  
\qquad \qquad \ \  
b_3 = -0.02656462, \\ 
&a_4 = -0.33369862.  
\qquad \qquad
b_4 = 0.06539614. \\ 
\end{split}
\label{eq:aba864_coeff}
\end{equation}

\section{Chaotic Properties of the Interpolating Lattice Models}
\label{Sec:Chaos_Indicators}

\begin{figure}
    \centering
    \includegraphics[trim={2cm 1cm 2cm 2cm},clip,scale=0.04]{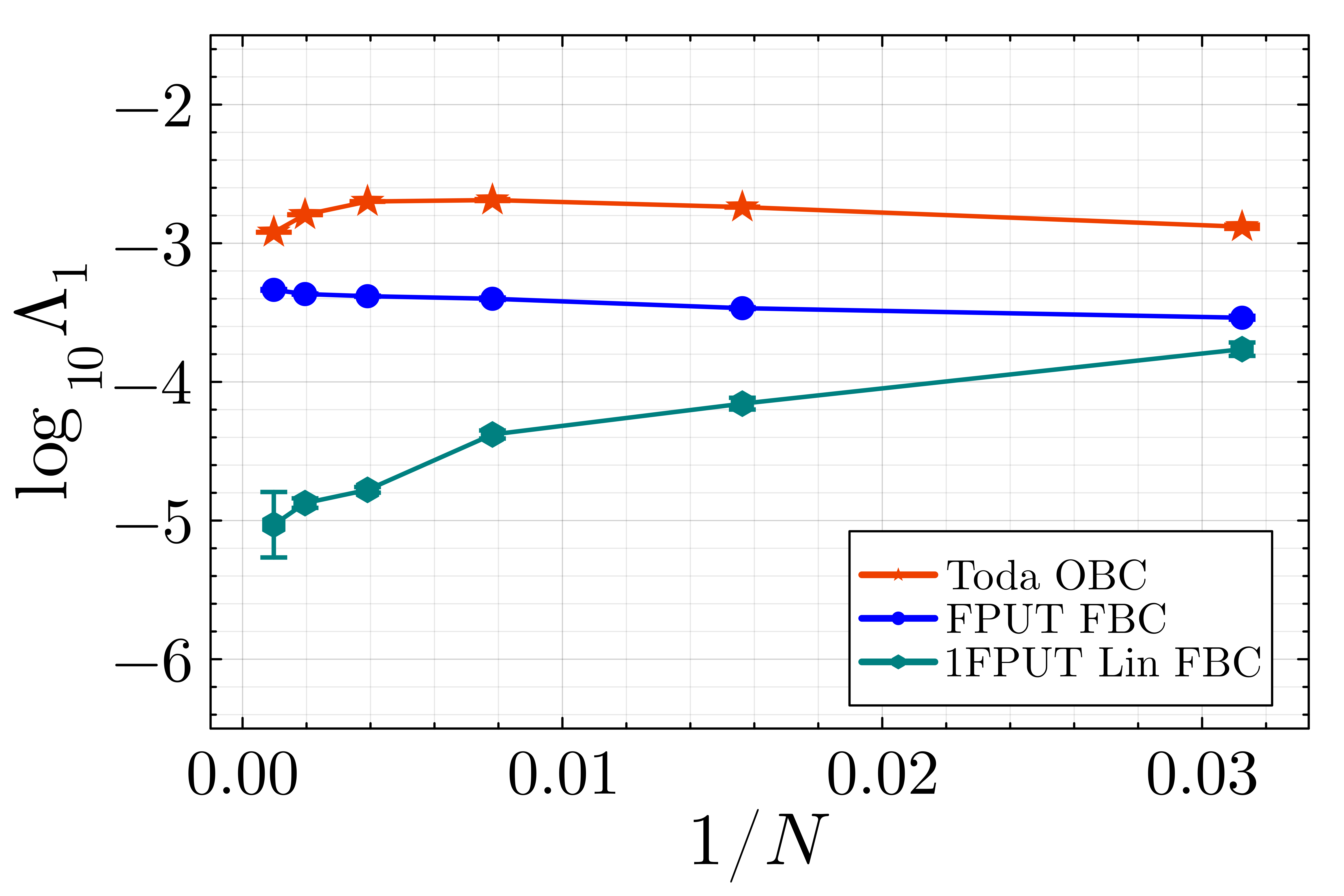}
    \caption{ We plot $\log_{10}\Lambda_1$, where $\Lambda_1$ denotes the maximal Lyapunov characteristic exponent (mLCE), as a function of inverse system size $1/N$ for three models: the fixed-boundary FPUT-$\alpha$ chain (blue circles), a linear chain with a single FPUT-$\alpha$ nonlinearity on the bond between sites $N/2$ and $N/2+1$ (teal hexagons), and the open-boundary Toda chain (red stars). In all three cases, the nonlinearity parameter is fixed to $\alpha = 0.25$. The initial conditions are randomly chosen at fixed energy density $E/N=0.1$, following the procedure used in Figs.~\ref{Fig:FPUT_FBC}, \ref{Fig:Mixed_FBC}, and \ref{Fig:Toda_MBC}. For the FPUT-$\alpha$ chain, $\Lambda_1$ shows almost no dependence on $N$. In contrast, for the other two models $\Lambda_1$ approaches a finite saturation value with increasing $N$, i.e., $\log_{10}\Lambda_1$ tends to a constant as $1/N \to 0$. }
    \label{Fig:LLE_NScaling}
\end{figure}

In this section, we calculate the Lyapunov spectrum (LS) for the following three scenarios:
\begin{enumerate}
    \item We first consider the FPUT-$\alpha$ potential with fixed boundary conditions, $V_{\textrm{FPUT-}\alpha,\textrm{FBC}}$, in the limit $\alpha \to 0$, where the system approaches the integrable linear chain since $V_{\textrm{FPUT-}\alpha,\textrm{FBC}}(\alpha=0)=V_{\textrm{lin},\textrm{FBC}}$. In Fig.~\ref{Fig:FPUT_FBC} we compute the LS for ten linearly spaced values of $\alpha$ in the range $0.1 \le \alpha \le 0.3$. Because the FPUT-$\alpha$ potential is unbounded from below, the parameter range must be restricted; for $\alpha \gtrsim 0.3$ the chain becomes unstable and undergoes breakdown \cite{FPUT_Breakdown}. 
    
    \item An interpolation between the fixed-boundary Toda and FPUT-$\alpha$ chains using the mixed potential defined in
    \begin{equation}
    \label{V_Mixed_FBC}
        V_\textrm{Mixed, FBC} = (1 - \varepsilon) V_\textrm{T,FBC} + \varepsilon V_\textrm{FPUT-$\alpha$, FBC},
    \end{equation}
    with $0<\varepsilon<1$. Here $\varepsilon=0$ corresponds to the integrable fixed-boundary Toda chain, while $\varepsilon=1$ yields the nonintegrable fixed-boundary FPUT-$\alpha$ chain. Figure~\ref{Fig:Mixed_FBC} shows the LS computed for  ten values of $\varepsilon$ between $0.1$ and $1.0$, chosen to be equally spaced on a logarithmic scale.
    
    \item The mixed-boundary Toda potential $V_{\textrm{T,Mixed}}$ (see Fig.~\ref{Fig:Toda_MBC}) defined in
    \begin{equation}
    \label{V_Toda_MBC}
        V_\textrm{Toda, Mixed} = V_\textrm{Toda, FBC} - \varepsilon \left[V_\textrm{Toda}\left(q_{1}\right) + V_\textrm{Toda}\left(- q_{N}\right)\right],
    \end{equation}
    considered in the limit $\varepsilon \to 0$. The case $\varepsilon=0$ corresponds to the integrable fixed-boundary Toda chain, while $\varepsilon=1$ yields the nonintegrable open-boundary Toda chain. In Fig.~\ref{Fig:Toda_MBC} we compute the LS for the same values of $\varepsilon$ used in Fig.~\ref{Fig:Mixed_FBC}.
\end{enumerate}

All Lyapunov characteristic exponent (LCE) calculations are performed using the $ABA864$ algorithm with parameters $\delta t = 0.32$, $T_{\mathrm{end}} = 10^{6}$, and energy density $E/N = 0.1$. The timestep and integration time are chosen such that the LS reaches clear saturation. In addition, the minimum values of $\alpha$ and $\varepsilon$ are selected to ensure that the LS converges within the integration window $t \le T_{\mathrm{end}}$. To quantify convergence, we estimate the uncertainties $\delta \Lambda_i$ from the standard deviation of the time-dependent exponents $\Lambda_i(t)$ over the window $t \in [T_{\mathrm{end}}/10, T_{\mathrm{end}}]$. Standard error-propagation is then used to obtain the uncertainties in $\log_{10}\Lambda_1$, $\overline{\Lambda}(\rho)$, and $\kappa$. 

\begin{figure}
    \centering
    \includegraphics[trim={2.5cm 1.5cm 2cm 1.5cm},clip,scale=0.4]{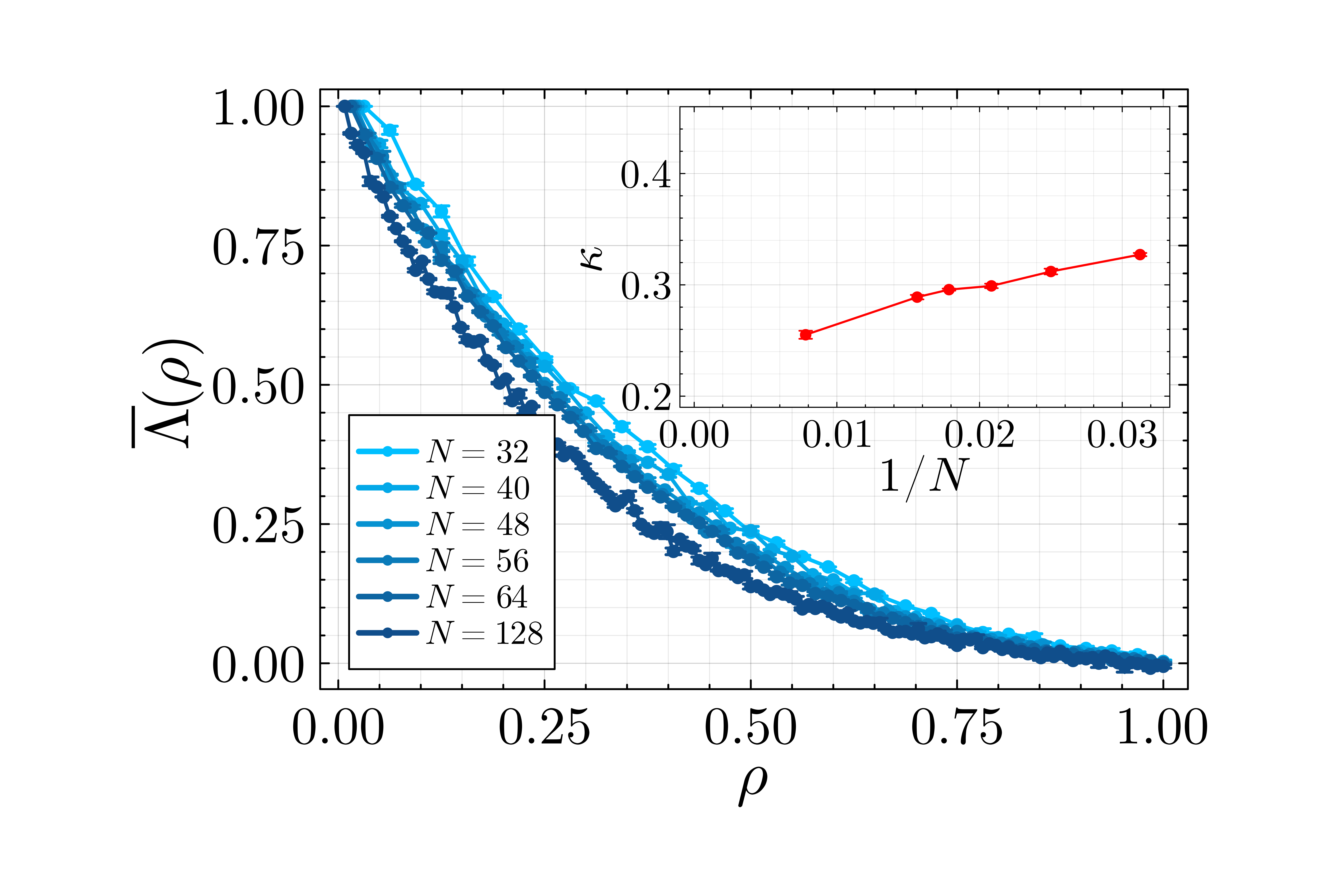}
    \caption{Rescaled Lyapunov spectra $\overline{\Lambda}(\rho) \equiv \Lambda_i/\Lambda_1$ as a function of $\rho = i/N$ $(i=1,\ldots,N)$ for open-boundary Toda chains at increasing system sizes. Results are obtained from random initial conditions at fixed energy density $E/N=0.1$ for $N=32,40,48,56,64,$ and $128$. The spectra exhibit negligible dependence on $N$. The inset shows the rescaled Kolmogorov-Sinai (KS) entropy $\kappa$ as a function of $1/N$, highlighting the convergence of the $\overline{\Lambda}(\rho)$ curves with increasing system size (cf. the (c) panels of Figs.~\ref{Fig:FPUT_FBC}, \ref{Fig:Mixed_FBC}, and \ref{Fig:Toda_MBC}). }
\label{Fig:Toda_OBC_LS_NScaling}
\end{figure}

It is well known that numerical simulations of integrable Hamiltonian systems with split-step symplectic integrators generically break integrability and can induce timestep- and integrator-dependent Trotter chaos \cite{AP,Patra}. Since integrable systems have vanishing LCEs, even a small spurious saturation value can be significant. In strongly chaotic systems such effects are typically negligible compared with the intrinsic LCEs. However, near integrability such protection is not guaranteed. We nevertheless find that our choice of $\delta t$, $T_{\mathrm{end}}$, and the fourth-order symplectic scheme $ABA864$ suppresses Trotter-induced artifacts to a negligible level \cite{AP}.

We now specify the initial conditions. Throughout, we use random initial conditions at fixed energy density $E/N \equiv h = 0.1$. All lattice displacements are set to zero, $q_j = 0$. The momenta are generated by first sampling $N$ random numbers $\mathfrak{p}_{0j}$ uniformly from $[0,1)$ and removing the mean, $\mathfrak{p}_j = \mathfrak{p}_{0j} - \overbar{\mathfrak{p}}_0$, where $\overbar{\mathfrak{p}}_0 = \left(\sum_{j=1}^{N} \mathfrak{p}_{0j}\right)/N$. The physical momenta are then obtained by rescaling \[p_{j} =  \left[\frac{h N}{H_\textrm{lattice}(\bm{q} = 0, \bm{p} = \bm{\mathfrak{p}})}\right]^{1/2}\mathfrak{p}_{j},\]
with $h=0.1$. This construction yields random initial conditions with the desired energy density irrespective of the form of $H_{\textrm{lattice}}$. We fix the random-number generator and seed to produce a single realization, which is used in all three scenarios shown in Figs.~\ref{Fig:FPUT_FBC}, \ref{Fig:Mixed_FBC}, and \ref{Fig:Toda_MBC}.

As emphasized in the Introduction, our goal is to probe the universality classes of weakly nonintegrable thermalization slowdown beyond the KAM regime. For the FPUT-$\alpha$ chain and related systems, this requires choosing an energy density $E/N$ above the stochasticity threshold $h_c$ \cite{FPUT_SST,FPUT_NLin_Res,Lichtenberg1,Lichtenberg2,FPUT_Breathers_Res_1}. For random initial conditions, Ref.~\cite{FPUT_Int_Dyn_03} estimates $h_c \propto N^{-2}$. The FPUT-$\alpha$ chain can also be viewed as a perturbation of the linear chain with effective strength $|\alpha|\sqrt{h}$ \cite{FPUT_Breathers_Res_2,FPUT_Int_Dyn_4}, suggesting that the stochasticity crossover occurs when $|\alpha|\sqrt{h}$ reaches a size-dependent threshold. Combining these considerations yields $h_c \sim C/(\alpha^2 N^2)$. Taking $C=\mathcal{O}(1)$ gives $h_c \approx 1.6\times10^{-2}$ for the parameters used here. Consistently, the analysis of $\Lambda_1$ in Ref.~\cite{FPUT_Int_Dyn_03} places the threshold in the range $10^{-3} < h_c < 10^{-2}$ (see Fig.~6 of \cite{FPUT_Int_Dyn_03}). These estimates indicate that our choice $E/N=0.1$ lies safely above the stochasticity threshold while remaining in the weakly nonintegrable regime.

To determine the network class of the three cases introduced above, we compute the positive part of the rescaled Lyapunov spectrum $\overline{\Lambda}(\rho) \equiv \Lambda_i/\Lambda_1$ with $\rho = i/N$ and $i=1,\ldots,N$, as defined below Eq.~\eqref{Lya_Spect_MET}. The rescaled Kolmogorov–Sinai (KS) entropy $\kappa$ is then obtained from Eq.~\eqref{KS_Entropy}. Figures~\ref{Fig:FPUT_FBC}, \ref{Fig:Mixed_FBC}, and \ref{Fig:Toda_MBC} show: (a) $\log$–$\log$ plots of $\Lambda_1$ versus the control parameters $\alpha$ or $\varepsilon$ governing the approach to integrability, (b) the curves $\overline{\Lambda}(\rho)$ versus $\rho$, and (c) the rescaled KS entropy $\kappa$ versus the same control parameter.

The three figures exhibit remarkably similar behavior. Panel (a) shows that $\Lambda_1$ decreases monotonically as the integrable limit is approached. In panel (b), the curves $\overline{\Lambda}(\rho)$ progressively collapse onto a common profile as $\alpha$ or $\varepsilon$ is reduced toward integrability. The limiting curves display a power-law decay of $\overline{\Lambda}(\rho)$ with $\rho$, consistent with the behavior reported in Ref.~\cite{Gabriel}. This convergence is further supported by panel (c), where the rescaled KS entropy $\kappa$ approaches a finite constant as the system approaches integrability. Since $\kappa$ is simply the area under the $\overline{\Lambda}(\rho)$–$\rho$ curve [Eq.~\eqref{KS_Entropy}], the saturation of $\kappa$ corroborates the collapse of the spectra. Together, these indicators point to a slowdown in thermalization through the long-range network (LRN) pathway in all three scenarios.

Recall from the Introduction that the LRN scenario arises when breaking integrability couples an extensive number of actions. In the first two cases -- approaching the fixed-boundary harmonic and Toda limits from the fixed-boundary FPUT-$\alpha$ chain -- the emergence of the LRN class is therefore natural. The actions of both the Toda and harmonic chains are highly nonlocal in lattice space, as shown in Eqs.~\eqref{Toda_Actions} and \eqref{Actions_Angles_SHO}. Consequently, even weak departures from these integrable limits couple many actions simultaneously, leading to the observed LRN thermalization slowdown.

More surprising is the appearance of the LRN class when approaching the fixed-boundary Toda chain from the open-boundary one. As seen from Eq.~\eqref{V_Toda_MBC}, the open-boundary Toda potential differs from the fixed-boundary case only by two boundary terms. One might therefore expect that breaking integrability through such a small number of local perturbations would produce chaos that vanishes in the thermodynamic limit. However, Fig.~\ref{Fig:LLE_NScaling} shows that $\Lambda_1$ instead saturates to a finite value as the system size $N$ increases for both models where integrability is broken locally. In addition, we compute the rescaled Lyapunov spectra for open-boundary Toda chains at increasing system sizes in Fig.~\ref{Fig:Toda_OBC_LS_NScaling}, starting from random initial conditions with fixed energy density $E/N=0.1$. These spectra exhibit only weak dependence on $N$.

This behavior can be understood by comparing the linear chain with a single FPUT-$\alpha$ nonlinearity to the full FPUT-$\alpha$ chain, as discussed in Appendix~\ref{App_SHO_1FPUT}. By counting the number of terms in the perturbation, one finds that the effective perturbation strength of a single FPUT-$\alpha$ defect is comparable to that of the fully nonlinear chain, owing to the nonlocal structure of the linear-chain actions. The same reasoning applies to the open-boundary Toda chain, explaining the emergence of the LRN class in that case as well.

\section{Summary and Outlook}
\label{Sec:Conclusion}

We conclude by summarizing our main results. We have investigated thermalization slowdown in three scenarios. In the first two, the integrable fixed-boundary harmonic and Toda chains are approached from the fixed-boundary FPUT-$\alpha$ chain, while in the third the fixed-boundary Toda chain is approached from the open-boundary Toda chain. In all cases, the scaling properties of the Lyapunov spectra (LS) demonstrate that the slowdown occurs through the long-range network (LRN) pathway.

One important scenario not addressed here is the approach to integrability in the fixed-boundary FPUT-$\alpha$ chain by decreasing the energy density. This problem is more subtle because the system admits two integrable limits -- the linear chain and the Toda chain -- and because the dynamics may enter a KAM regime. Understanding how the KAM regime affects the two network classes of thermalization slowdown would therefore be an important direction for future work.

Another open question concerns the role of the nonuniqueness of action–angle representations in determining the scaling properties of the LS. For example, Ref.~\cite{Cyprus} shows that the open-boundary Toda lattice with bare exponential interactions is superintegrable: a chain with $N$ sites possesses $(2N-1)$ independent integrals of motion. As a consequence, the actions obtained from the Hamilton–Jacobi equations are not unique in this model. It would be interesting to determine how the LS behaves when this integrable limit is approached from a nonintegrable system such as the open-boundary Toda chain.

A related direction is to explore networks associated with integrable chains that host flat bands \cite{Flat-Band_1, Flat-Band_2, Flat-Band_3}. A flat-band normal-mode spectrum implies strong degeneracies in action space. In contrast to the partial degeneracy of the open-boundary Toda lattice, the degeneracy here appears to be maximal. Whether such systems give rise to a distinct network class of thermalization slowdown remains an intriguing open question.

\appendix 
\section{FPUT-$\alpha$ Lattice in Terms of Harmonic Chain Actions and Angles}
\label{App_SHO_1FPUT}

Consider the classical harmonic chain Hamiltonian with fixed boundary condition
\begin{equation}
    H_\textrm{lin, FBC} = \sum_{j = 0}^{N} \left[ \frac{p_j^2}{2} + \frac{1}{2}\left(q_{j+1} - q_{j}\right)^2 \right],
    \label{H_SHO}
\end{equation}
where we have $q_0 = q_{N+1} = p_{0} = p_{N+1} = 0$. We introduce the normal mode positions and momenta for this Hamiltonian as
\begin{equation}
\begin{aligned}
    \begin{pmatrix}
        q_{j} \\ p_{j}
    \end{pmatrix} = \sqrt{\frac{2}{N+1}}\sum_{k=1}^{N} 
    \begin{pmatrix}
        Q_{k} \\ P_{k}
    \end{pmatrix} \sin{\left( \frac{\pi kj}{N+1} \right)} \coloneqq \sqrt{\frac{2}{N+1}}\sum_{k=1}^{N}\begin{pmatrix}
        Q_{k} \\ P_{k}
    \end{pmatrix} \sin{\left( 2\alpha_k j \right)},
\end{aligned}
\label{Normal_Modes_SHO}
\end{equation}
where the variable 
\begin{equation}
    \alpha_{k} = \frac{\pi k}{2(N+1)}
    \label{defn_alpha_q}
\end{equation}
is introduced for later convenience. The normal mode variables \eqref{Normal_Modes_SHO} can then be related to the actions and angles of the harmonic chain as
\begin{equation}
    Q_{k} = \sqrt{2J_{k}} \sin{\theta_{k}}, \quad P_{k} = \omega_{k}\sqrt{2J_{k}} \cos{\theta_{k}},  
    \label{Actions_Angles_SHO}
\end{equation}
where 
\begin{equation}
    \omega_{k} = 2\sin{\left(\frac{\pi k}{2(N+1)}\right)} \coloneqq 2\sin{\left(\alpha_{k}\right)}.
    \label{SHO_Dispersion}
\end{equation}
Equation \eqref{H_SHO} can then be rewritten in terms of these new variables as 
\begin{equation}
    H_\textrm{lin, FBC} = \sum_{k = 1}^{N} \frac{P_k^2 + \omega_{k}^2Q_k^2}{2} = \sum_{k = 1}^{N} \omega_{k}^2 J_k. 
    \label{H_SHO_NM_AA}
\end{equation}
We want to treat \(H_\textrm{FPUT-$\alpha$, FBC}\) as a perturbation over \(H_\textrm{lin, FBC}\). To do so, we recast \(H_\textrm{FPUT-$\alpha$, FBC}\) in terms of \(J_{k}\) and \(\theta_k\) from Eq.\ \eqref{Actions_Angles_SHO}. We start by rewriting the nonlinear term in the potential as 
\begin{equation}
    \Delta V_\textrm{FPUT-$\alpha$, FBC} = \sum_{j = 0}^{N} \frac{\alpha}{3}\left(q_{j+1} - q_{j}\right)^3 \equiv \sum_{j = 0}^{N} \frac{\alpha}{3}\left( \Delta q_{j} \right)^3. 
    \label{Pert_FPUT}
\end{equation}
We write \(\Delta q_{i}\) in terms of the coordinates of the normal mode as
\begin{equation}
    \begin{aligned}
        \Delta q_{j} = \sqrt{\frac{2}{N+1}} \sum_{k=1}^{N} Q_k\left[\sin{(2\alpha_{k}(j+1))} - \sin{(2\alpha_{k}j)}\right] = \sum_{k=1}^{N} c_k \cos{[(2j+1)\alpha_{k}]}, 
    \end{aligned} 
    \label{Del_q}
\end{equation}
where 
\begin{equation}
    c_k = 2\sqrt{\frac{2}{N+1}}Q_k \sin{\alpha_{k}}.
    \label{c_q_defn}
\end{equation}
In the final line of Eq.\ \eqref{Del_q} we have used
\begin{equation}
    \sin{A} - \sin{B} = 2\cos{\left(\frac{A+B}{2}\right)} \sin{\left(\frac{A-B}{2}\right)}.
\end{equation}

Cubing both sides of Eq.\ \eqref{Del_q}, we obtain 
\begin{equation}
    \begin{aligned}
        (\Delta q_{j})^3 &= \sum_{k_1, k_2, k_3} c_{k_1} c_{k_2} c_{k_3} \prod_{r = 1}^{3} \cos{[(2j+1)\alpha_{k_r}]} \\
        &= \frac{1}{4} \sum_{k_1, k_2, k_3} c_{k_1} c_{k_2} c_{k_3} \sum_{\sigma_1, \sigma_2 \in \{\pm1\}} \cos{[(2j+1)\Lambda(\bm{k}, \bm{\sigma})]} \\
        &= \frac{1}{\sqrt{2}} \frac{1}{\left(N+1\right)^{3/2}} \sum_{k_1, k_2, k_3}Q_{k_{1}} Q_{k_{2}} Q_{k_{3}} \omega_{k_1} \omega_{k_2} \omega_{k_3}  \sum_{\sigma_1, \sigma_2 \in \{\pm1\}} \cos{[(2j+1)\Lambda(\bm{k}, \bm{\sigma})]}
    \end{aligned} 
    \label{Del_q^3}
\end{equation}
where 
\begin{equation}
    \Lambda(\bm{k}, \bm{\sigma}) \coloneqq \alpha_{k_1} + \sigma_1 \alpha_{k_2} + \sigma_2 \alpha_{k_3}.
    \label{Lambda_defn}
\end{equation}
In the last line of Eq.\ \eqref{Del_q^3}, we have used
\begin{equation}
\begin{aligned}
    \cos{A}\cos{B}\cos{C} = \frac{1}{4} \left[ \cos{\left( A+B+C \right)} + \cos{\left( A+B-C \right)} + \cos{\left( A-B+C \right)} + \cos{\left( A-B-C \right)} \right].
\end{aligned}
\label{Cosine_Identity}
\end{equation}
In order to obtain an expression for $\Delta V_\textrm{FPUT-$\alpha$, FBC}$, we need to evaluate $\sum_{j=0}^{N} (\Delta q_{j})^3$. To that end, we note that
\begin{equation}
\begin{aligned}
    \mathcal{S}(N, \bm{k}, \bm{\sigma}) &=\sum_{j=0}^{N}\cos{[(2j+1)\Lambda(\bm{k}, \bm{\sigma})]} \\
    &= \mathrm{Re} \left[ \sum_{j=0}^{N} e^{i(2j+1)\Lambda(\bm{k}, \bm{\sigma})} \right]  \\
    &= \mathrm{Re} \left[ e^{i\Lambda(\bm{k}, \bm{\sigma})} \frac{1 - e^{2i(N+1)\Lambda(\bm{k}, \bm{\sigma})}}{1 - e^{2i\Lambda(\bm{k}, \bm{\sigma})}}\right] \\
    &= \frac{ \sin{[2(N+1)\Lambda(\bm{k}, \bm{\sigma})]} }{ 2\sin{[\Lambda(\bm{k}, \bm{\sigma})]} } \\
    &=\sin{[ \pi \underbrace{(k_1 + \sigma_1 k_2 + \sigma_2 k_3)}_{\mathrm{ integer}:= \mathbb{k}(\bm{k}, \bm{\sigma})} ]}/ 2\sin{[\Lambda(\bm{k}, \bm{\sigma})]}. 
\end{aligned}
\label{Cosine_Sum}
\end{equation}
As a result, the above sum vanishes unless \(\Lambda(\bm{k}, \bm{\sigma}) = n\pi\), where \(n\) is an integer. Because the \(\cos\) function is even and \(\pi\)-antiperiodic, restricting \(\Lambda(\bm{k}, \bm{\sigma})\) into the domain $[0, \pi]$ is enough to obtain all the possible values of \(\mathcal{S}(N, \bm{k}, \bm{\sigma})\). Using this, we have
\begin{equation}
    \begin{aligned}
        \mathcal{S}(N, \bm{k}, \bm{\sigma}) = \begin{cases}
            N+1, \quad &\textrm{if }\Lambda(\bm{k}, \bm{\sigma}) = 0, \\
            -(N+1), \quad &\textrm{if }\Lambda(\bm{k}, \bm{\sigma}) = \pi, \\
            0, \quad &\textrm{otherwise.}
        \end{cases}
    \end{aligned}
    \label{Sum_Resonance_Condition}
\end{equation}
In terms of $\bm{k}$ and $\bm{\sigma}$, we translate the resonance conditions of \eqref{Sum_Resonance_Condition} as follows:
\begin{subequations}
    \begin{align}
        &\Lambda(\bm{k}, \bm{\sigma}) = 0 \Longleftrightarrow \mathbb{k}(\bm{k}, \bm{\sigma}) = 0 \Longleftrightarrow k_3 = k_1 + k_2, k_2 = k_3 + k_1, k_1 = k_2 + k_3,  \label{Sum_Diff}\\
        &\Lambda(\bm{k}, \bm{\sigma}) = \pi \Longleftrightarrow \mathbb{k}(\bm{k}, \bm{\sigma}) = 2(N+1) \Longleftrightarrow k_1 + k_2 + k_3 = 2(N+1). \label{Folded}
    \end{align}
    \label{Discrete_Mom_Consv}
\end{subequations}
Equation \eqref{Discrete_Mom_Consv} lists all the different choices of $\bm{\sigma}$ for a given $\bm{k}$, for which $\mathcal{S}(N, \bm{k}, \bm{\sigma}) \neq 0$. Putting Eqs.\ \eqref{Pert_FPUT}, \eqref{Del_q^3}, \eqref{Sum_Resonance_Condition}, and \eqref{Discrete_Mom_Consv} together, we obtain
\begin{equation}
    \begin{aligned}
        &\Delta V_\textrm{FPUT-$\alpha$, FBC} \\
        &= \frac{\alpha}{12} \sum_{\bm{k}, \bm{\sigma}} c_{k_1} c_{k_2} c_{k_3} \underbrace{\sum_{j=0}^{N} \cos{[(2j+1)\Lambda(\bm{k}, \bm{\sigma})]}}_{\mathcal{S}(N, \bm{k}, \bm{\sigma})} \\
        &= \frac{\alpha}{3} \frac{2^{5/2}}{\sqrt{N+1}} \sum_{k_1, k_2, k_3 = 1}^{N} Q_{k_1} Q_{k_2} Q_{k_3} \sin{\left(\alpha_{k_1}\right)} \sin{\left(\alpha_{k_2}\right)} \sin{\left(\alpha_{k_3}\right)} \left[ \delta_{k_3, k_1+k_2} + \delta_{k_2, k_3+k_1} + \delta_{k_1, k_2+k_3} - \delta_{k_1 + k_2 + k_3, 2(N+1)} \right] \\
        &= \frac{\alpha}{3\sqrt{2(N+1)}} \sum_{k_1, k_2, k_3 = 1}^{N} Q_{k_1} Q_{k_2} Q_{k_3} \omega_{k_1} \omega_{k_2} \omega_{k_3} \left[ \delta_{k_3, k_1+k_2} + \delta_{k_2, k_3+k_1} + \delta_{k_1, k_2+k_3} - \delta_{k_1 + k_2 + k_3, 2(N+1)} \right].
    \end{aligned} 
    \label{Pert_FPUT_Final_Form}
\end{equation}
We compare the following two perturbations: 
\begin{enumerate}
    \item $\Delta V_\textrm{FPUT-$\alpha$, FBC}$: Sum of cubic FPUT-$\alpha$ nonlinearities at each bond.
    \item $\Delta V^{j}_\textrm{FPUT-$\alpha$, FBC}$: One single cubic FPUT-$\alpha$ nonlinearity at the $j$th bond.  
\end{enumerate}
The perturbations can be written as  
\begin{equation}
\label{Compare_Pert_Full_vs_One_FPUT_Nonlin}
    \begin{aligned}
        &\Delta V_\textrm{FPUT-$\alpha$, FBC} \\
        &= \frac{\alpha}{3\sqrt{2(N+1)}} \sum_{k_1, k_2, k_3 = 1}^{N} Q_{k_1} Q_{k_2} Q_{k_3} \omega_{k_1} \omega_{k_2} \omega_{k_3} \left[ \delta_{k_3, k_1+k_2} + \delta_{k_2, k_3+k_1} + \delta_{k_1, k_2+k_3} - \delta_{k_1 + k_2 + k_3, 2(N+1)} \right], \\ \\
        &\Delta V^{j}_\textrm{FPUT-$\alpha$, FBC} \\
        &= \frac{\alpha}{3\sqrt{2}} \frac{1}{\left(N+1\right)^{3/2}} \sum_{k_1, k_2, k_3}Q_{k_{1}} Q_{k_{2}} Q_{k_{3}} \omega_{k_1} \omega_{k_2} \omega_{k_3} \sum_{\sigma_1, \sigma_2 \in \{\pm1\}} \cos{[(2j+1)\Lambda(\bm{k}, \bm{\sigma})]}
    \end{aligned} 
\end{equation}
In the expression for $\Delta V_\textrm{FPUT-$\alpha$, FBC}$, the $\delta$ functions gets rid of a summation. The other two summations provide us with 
\begin{equation}
\label{Estimate_Full_FPUT_Nonlinearity_Pert}
\begin{aligned}
    \Delta V_\textrm{FPUT-$\alpha$, FBC} &\sim \mathcal{O}\left(N^{-1/2}\right) \times \mathcal{O}\left(N\right) \times \mathcal{O}\left(N\right) \\ 
    &= \mathcal{O}\left(N^{3/2}\right).
\end{aligned}
\end{equation}
Using a similar argument and neglecting the $\cos$ term, we obtain 
\begin{equation}
\label{Estimate_One_FPUT_Nonlinearity_Pert}
\begin{aligned}
    \Delta V^{j}_\textrm{FPUT-$\alpha$, FBC} &\sim \mathcal{O}\left(N^{-3/2}\right) \times \mathcal{O}\left(N\right) \times \mathcal{O}\left(N\right) \times \mathcal{O}\left(N\right) \\ 
    &= \mathcal{O}\left(N^{3/2}\right).
\end{aligned}
\end{equation}
Even though the strengths of the perturbations are comparable, we observe that $\Lambda_{1}$ decreases as $\mathcal{O}\left(N^{-1}\right)$ for $\Delta V^{j}_\textrm{FPUT-$\alpha$, FBC}$, whereas it saturates for $\Delta V_\textrm{FPUT-$\alpha$, FBC}$.

\begin{acknowledgments}
The authors acknowledge financial support from the Institute for Basic Science (IBS) in the Republic of Korea through Project No.
IBS-R024-D1 and IBS-R041-D1-2026-a00. Additionally, one of us (A.P.) was partially supported by the National Research Foundation of Korea (NRF) through a grant funded by the Korean government (MSIT) (No. RS-2025-02315685).
\end{acknowledgments}

\section*{Statements and Declarations}

\textbf{Conflict of Interest:} On behalf of all authors, the corresponding author states that there is no conflict of interest.

\textbf{Data Availability:} The data supporting the findings of this study are available from the corresponding author upon reasonable request.

\end{document}